\documentclass[twocolumn,showpacs,preprintnumbers,amsmath,amssymb]{revtex4}
\usepackage{amssymb}
\usepackage{amsfonts}
\usepackage{mathrsfs}
\usepackage{amsmath}
\usepackage{graphicx}
\usepackage{dcolumn}
\usepackage{bm}
\usepackage{hyperref}

\begin{document}
\title{Passive faraday mirror attack in practical two-way quantum key distribution system}
\author{Shi-Hai Sun, Mu-Sheng Jiang, Lin-Mei Liang\footnote{Email:nmliang@nudt.edu.cn}}
\affiliation{Department of Physics, National University of Defense
Technology, Changsha 410073, P.R.China}
\begin{abstract}
The faraday mirror (FM) plays a very important role in maintaining the stability of two way \emph{plug-and-play} quantum key distribution (QKD) system. However, the practical FM is imperfect, which will not only introduce additional quantum bit error rate (QBER) but also leave a loophole for Eve to spy the secret key. In this paper, we propose a passive faraday mirror attack in two way QKD system based on the imperfection of FM. Our analysis shows that, if the FM is imperfect, the dimension of Hilbert space spanned by the four states sent by Alice is three instead of two. Thus Eve can distinguish these states with a set of POVM operators belonging to three dimension space, which will reduce the QBER induced by her attack. Furthermore, a relationship between the degree of the imperfection of FM and the transmittance of the practical QKD system is obtained. The results show that, the probability that Eve loads her attack successfully depends on the degree of the imperfection of FM rapidly, but the QBER induced by Eve's attack changes with the degree of the imperfection of FM slightly.
\end{abstract}


\maketitle
\section{\label{sec:Introd}INTRODUCTION}
Quantum key distribution (QKD) \cite{Bennett} admits two remote parties, known as Alice and Bob, to share unconditional secret key, even the eavesdropper (Eve) has ultimate power admitted by the quantum mechanics. Although the unconditional security have been proved for both the ideal system \cite{Lo,Shor} and the practical system \cite{GLLP,Inamori} in past years, some assumptions are set to limit Eve's attack strategy or to ignore some imperfections existed in the practical QKD system. Generally speaking, the practical QKD system is imperfect. Any deviation between the standard security analysis and the practical QKD system will leave a loophole for Eve to obtain more information. In the worst case, Eve can exploit all these imperfections together to maximize her information about the secret key. Thus it is important to do research on the practical QKD system carefully and close these loopholes to guarantee the unconditional security of key. In fact, some potential attacks using the imperfection of a practical QKD system have been discovered, for example, timing side channel attack \cite{Lamas}, faked states attack \cite{Makarov}, blinding attack \cite{Lydersen}, Trojan-horse attacks \cite{Gisin}, time-shifted attack \cite{Zhao,MAS}, phase-remapping attack \cite{Fung,Xu}. Therefore, when the QKD system is used in the practical situation, the legitimate parties should consider the potential attack according to any imperfection existed in the practical system and find defense strategies against them.

In all the practical QKD system based on long distance fiber, the major difficulty is to maintain the stability and compensate the birefringence of fiber. In order to resolve this problem, Muller \emph{et al.} proposed an interesting two-way \emph{plug-and-play} scheme \cite{Muller}, which can compensate the birefringence automatically. In this system, Bob sends a strong reference pulse to Alice. Then Alice encodes her information to the reference pulse, attenuates it to single photon level, and sends it back to Bob. Since the pulse travels back and forth in the quantum channel, the birefringence is compensated automatically. However, since Alice admits the pulse go in and go out of her zone, it will leave a backdoor for Eve to implement variable Trojan-horse attack \cite{Gisin,Zhao,MAS,Fung,Xu}.

In this paper, we propose a passive faraday mirror (PFM) attack in two way \emph{plug-and-play} QKD system based on the imperfection of faraday mirror (FM) which plays a very important role in compensating the birefringence of fiber. Our results show that, for the BB84 protocol \cite{Bennett}, when the FM deviates from the ideal situation, the dimension of Hilbert space spanned by the states sent by Alice is three instead of two. Thus it will give Eve more information to spy the secret key. When the legitimate parties are unaware of our attack, unconditional security of the generated key must be compromised. Thus, in practical situation, it is very important for Alice and Bob to consider our attack when they judge whether the \emph{plug-and-play} QKD system is secure or not.

In the following, we first, in Sec.\ref{sec:fm} and Sec.\ref{sec:attack}, introduce the imperfection of FM and analyze a PFM attack based on this imperfection. In Sec.\ref{sec:sim}, we find the minimal QBER between Alice and Bob induced by Eve, when she uses an optimal and suboptimal POVM measurement strategy to implement the intercept-and-resend attack. In Sec.\ref{sec:con}, we give a brief conclusion of this paper.

\section{\label{sec:fm}The imperfection of faraday mirror}
In this section, we first introduce the two way \emph{plug-and-play} system briefly, and show why the FM can be used to compensate the birefringence of fiber. Then we show how the imperfection of FM can be used by Eve to spy the secret key.
\subsection{Plug-and-play QKD system}
\begin{figure}
\scalebox{1}{\includegraphics[width=\columnwidth]{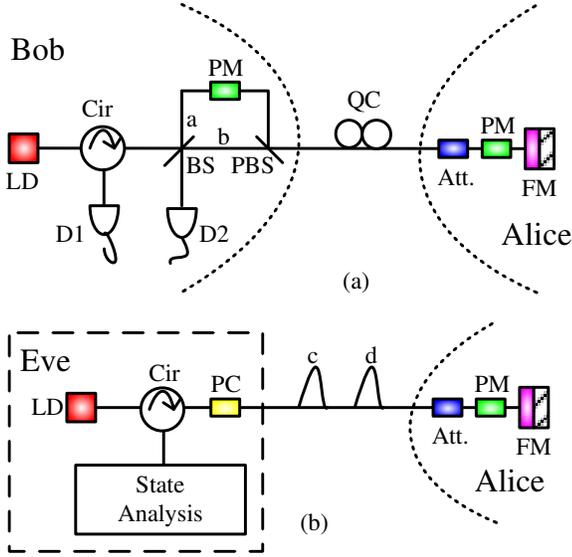}}
\caption{\label{fig:1}(Color online)The simple diagram of plug-and-play system \cite{Muller} and Eve's attack. LD:laser diode, Cir: circulator, BS: beam splitter, PBS: polarization beam splitter, Att.: attenuator, PM: phase modulator, FM: faraday mirror, QC: quantum channel, D1 and D2: single photon detector, PC: polarization controller. Part(a) shows plug-and-play system without Eve. Part(b) shows Eve's PFM attack. In the diagram, we only show how Eve obtains the information of Alice. \emph{c} and \emph{d} are two time modes sent to Alice by Eve and we assume only mode c is modulated by Alice in this paper.}
\end{figure}
A simple diagram of \emph{plug-and-play} system \cite{Muller} without Eve is shown in Fig.\ref{fig:1}(a). Bob sends a strong reference pulse to Alice, which is horizontally polarized. The pulse will be divided equally into two parts by a beam splitter (BS), noted as \emph{a} and \emph{b}. A polarization controller (not shown in Fig.\ref{fig:1}) is used to change the polarization of \emph{b} to guarantee it can pass the polarization beam splitter (PBS) totally. Generally speaking, due to the birefringence of fiber, the polarization of \emph{a} and \emph{b} are random, when they arrive at Alice's zone sequentially. However, a FM can be used to compensate the birefringence of channel automatically. When \emph{a} and \emph{b} return Bob's zone, their polarization are orthogonal to that of their initial state. Then they will travel the other path and interfere in BS. Therefore, FM plays an important role in compensating the birefringence of fiber. Now, we show why the FM can do this.

The FM is a combination of a $\theta$ faraday rotator and an ordinary mirror. In ideal situation, $\theta=45^\circ$ and the Jones matrix of FM can be written as:
\begin{equation}
\label{fm_per}
\begin{split}
FM(45^\circ)&=\frac{1}{\sqrt{2}}\begin{bmatrix}1&1\\-1&1\end{bmatrix}
\begin{bmatrix}1 &0\\0&-1\end{bmatrix}
\frac{1}{\sqrt{2}}\begin{bmatrix}1&-1\\1&1\end{bmatrix}\\
&=-\begin{bmatrix}0&1\\1&0\end{bmatrix}.
\end{split}
\end{equation}

Thus the polarization of the outgoing state is always orthogonal to that of the incoming state, regardless of the input polarization state. It is easy to prove that for any birefringence medium, the following equation always holds, that is:
\begin{equation}
\label{fm_com}
T(-\theta')\cdot FM(45^\circ)\cdot T(\theta')=e^{i(\varphi_o+\varphi_e)}FM(45^\circ),
\end{equation}
where $T(\theta')$ and $T(-\theta')$ are the Jones matrices of birefringence medium when the photon travels forward and backward the quantum channel, which are given by:
\begin{equation}
\begin{split}
T(\pm\theta')&=\begin{bmatrix}cos(\theta') &\mp sin(\theta')\\ \pm sin(\theta') &cos(\theta') \end{bmatrix}\begin{bmatrix}exp(i\varphi_o) &0\\0 &exp(i\varphi_e)\end{bmatrix}\\
&\times\begin{bmatrix}cos(\theta') &\pm sin(\theta')\\ \mp sin(\theta') &cos(\theta') \end{bmatrix}
\end{split},
\end{equation}
where $\varphi_o$, $\varphi_e$ are the propagation phases of ordinary and extraordinary rays and $\theta'$ is the rotation angle between the reference basis and the eigenmode basis of the birefringence medium. Eq.\eqref{fm_com} shows clearly that, in the ideal situation, the \emph{plug-and-play} system can compensate the birefringence of medium automatically.

Here, we remark that although the \emph{plug-and-play} system will suffer from the ``untrusted source" problem in which the source incoming Alice's zone is controlled by Eve totally, the security has been rigorously proved in a few recent works \cite{Zhao10, Zhao08, Peng08, Xu10, Moroder09, Wang09}. Thus, this problem is not considered in this paper and the additional setups for Alice to monitor the ``untrusted source" are also not shown in Fig.\ref{fig:1}.

\subsection{PFM attack}
In the discussion above, we have shown that, in the ideal situation, FM can be used to compensate the birefringence of fiber automatically. However, in practical case, the angle of the faraday rotator in FM is not exact $45^\circ$. Then Eq.\eqref{fm_per} is not valid and the Jones matrix of a practical FM should be rewritten as:
\begin{equation}
\begin{split}
FM(\theta)&=\begin{bmatrix}cos(\theta)&sin(\theta)\\-sin(\theta)&cos(\theta)\end{bmatrix}
\begin{bmatrix}1 &0\\0&-1\end{bmatrix}
\begin{bmatrix}cos(\theta)&-sin(\theta)\\sin(\theta)&cos(\theta)\end{bmatrix}\\
&=\begin{bmatrix}cos(2\theta)&-sin(2\theta)\\-sin(2\theta)&-cos(2\theta)\end{bmatrix}\\
&=-\begin{bmatrix}sin(2\varepsilon)&cos(2\varepsilon)\\cos(2\varepsilon)&-sin(2\varepsilon)\end{bmatrix}\equiv FM(\varepsilon),
\end{split}
\end{equation}
where $\theta=\pi/4+\varepsilon$ is the angle of faraday rotator in a practical FM. Generally speaking, $\varepsilon$ is small. For example, in the center wavelengths 1550nm and 1310nm, the maximal rotation angle tolerance is $1^\circ$ (at $23^\circ C$ ) for the popular FM produced by \emph{Newport} \cite{Newprot} and \emph{General Photonics} \cite{GP}. Thus, in this paper, we only consider the case that $|\varepsilon|\leq 1^{\circ}$.

When FM is imperfect, the birefringence of fiber can not be compensated totally and additional QBER will be induced. However, the additional QBER is just the minor bug of the practical FM, since $\varepsilon$ is very small. The major one is that the imperfection of FM will leave a loophole for Eve to spy the secret key. In the following, we show how Eve can use this imperfection, which is called PFM attack in this paper.

The PFM attack is shown in part(b) of Fig.\ref{fig:1}. In the diagram, we only draw the major part of Eve's attack that how Eve probes Alice's information. In order to do this, Eve sends two time modes \emph{c} and \emph{d} to Alice. Note that the two modes should be coherent , which can be obtained by splitting a pulse with a BS like the generation of mode \emph{a} and \emph{b} sent by Bob. The polarization of the two modes should also be the same, which is controlled by Eve's polarization controller(PC). We assume the polarization state of photon that sent to Alice by Eve is given by:
\begin{equation}
in=\begin{bmatrix} \alpha & \beta \end{bmatrix}^{T}.
\end{equation}
Note that the polarization of incoming state is controlled by Eve totally, thus $\alpha$ and $\beta$ are any complex number that satisfy $|\alpha|^2+|\beta|^2=1$.  Simply, we only consider the the special case that Eve sets $\alpha=1$ and $\beta=0$ in this paper.

It is easy to prove that the Jones vectors of output polarization state for the two time modes can be written as:
\begin{eqnarray}
\label{out_c}
\begin{split}
out_c&=\begin{bmatrix}e^{ik\delta}&0\\0&1\end{bmatrix}\cdot FM(\varepsilon)
\cdot\begin{bmatrix}e^{ik\delta}&0\\0&1\end{bmatrix}\cdot in\\
&=-e^{ik\delta}\begin{bmatrix}sin(2\varepsilon)e^{ik\delta}\\cos(2\varepsilon)\end{bmatrix}\\
\end{split}\\
\label{out_d}
out_d=FM(\varepsilon)\cdot in=-\begin{bmatrix}sin(2\varepsilon)\\cos(2\varepsilon)\end{bmatrix}
\end{eqnarray}
where $out_c$ and $out_d$ are the Jones vectors of mode \emph{c} and mode \emph{d}. Here we assume that only \emph{c} is modulated by Alice. $k=0,1,2,3$ are the indices of the four states modulated by Alice. $\delta$ is the phase difference between states, in the standard BB84 scheme, $\delta=\pi/2$, but if Eve combines our attack with the phase-remapping attack \cite{Fung,Xu}, $\delta \in [0,\pi/2]$. Note that, in Eq.\eqref{out_c} and Eq.\eqref{out_d}, we assume that Eve can compensate perfectly the birefringence of Alice's zone by controlling her polarization controller. Although it is still challenging as the birefringence at Alice's side fluctuates, we can ignore this fluctuation, since the interval that Eve's photon incoming and outgoing Alice's zone is much smaller than the time for a remarkable fluctuation of the birefringence.

Therefore, when FM is imperfect, the states sent by Alice are not the standard BB84 state which are noted as $|\phi_k\rangle=(e^{ik\pi/2}|c\rangle+|d\rangle)/\sqrt{2}$, but four new states that can be written as:
\begin{equation}
\label{phi1}
\begin{split}
|\Phi_k\rangle &=\frac{1}{\sqrt{2}}\{sin(2\varepsilon)e^{i2k\delta}|cH\rangle+cos(2\varepsilon)e^{ik\delta}|cV\rangle\\
&+sin(2\varepsilon)|dH\rangle+cos(2\varepsilon)|dV\rangle\}.
\end{split}
\end{equation}
It is easy to check that, when $\varepsilon\neq 0$, the dimension of Hilbert space spanned by the four new states of Eq.\eqref{phi1} is three. In order to show it clearly, we let $|H\rangle=cos(2\varepsilon)|H'\rangle+sin(2\varepsilon)|V'\rangle$ and $|V\rangle=-sin(2\varepsilon)|H'\rangle+cos(2\varepsilon)|V'\rangle$, and note $|cH'\rangle=|e_0\rangle$, $|cV'\rangle=|e_1\rangle$, $|dV'\rangle=|e_2\rangle$. Note that $\varepsilon$ can be known by Eve exactly, thus the transformation from $\{H,V\}$ to $\{H',V'\}$ can be implemented by Eve. Then Eq.(8) can be rewritten as:
\begin{equation}
\label{phi}
\begin{split}
|\Phi_k\rangle&=\frac{1}{\sqrt{2}}\{sin(2\varepsilon)cos(2\varepsilon)(e^{i2k\delta}-e^{ik\delta})|e_0\rangle \\ &+[sin^2(2\varepsilon)e^{i2k\delta}+cos^2(2\varepsilon)e^{ik\delta}]|e_1\rangle+ |e_2\rangle\}.
\end{split}
\end{equation}

Eq.\eqref{phi1} and Eq.\eqref{phi} show clearly that the information encoded by Alice appears not only in the time mode (\emph{c} and \emph{d}), but also in the polarization mode (\emph{H} and \emph{V}). Obviously, Eve can obtain more information from the four new states than the four standard BB84 states. Thus if Alice and Bob are unaware of the imperfection of FM, it will leave a loophole for Eve to spy the information encoded by Alice. Here, we remark that, although the true states sent by Alice is shown by Eq.\eqref{phi1}, Eve can not get more information from Eq.\eqref{phi1} than Eq.\eqref{phi}, since the transformation from Eq.\eqref{phi1} to Eq.\eqref{phi} is just one unitary operation which can not give her more information. Thus, in the following, we use Eq.\eqref{phi} but not Eq.\eqref{phi1} to continue our discussion.

\section{\label{sec:attack}PFM attack based on an intercept-and-resend attack}
When FM is imperfect, the states sent by Alice are not the standard BB84 states but the states shown by Eq.\eqref{phi}. Thus, Eve can use the operators belonging to 3-dimension Hilbert space to measure the four states. Note that, for the case that FM is perfect, the operators that can be used by Eve should be limited in 2-dimension Hilbert space. Generally speaking, when a new imperfection of system is found by Eve, she can combine all imperfections of the system and attack strategies to maximize her information of the key. However, in this paper, we only consider the intercept-and-resend attack, which shows clearly that the generated key will be compromised due to the imperfection of FM.

We consider the following attack, Eve intercepts each pulse returned from Alice's zone and measures it with five POVM operators $\{M_{vac},M_i|i=0,1,2,3\}$ which satisfy the condition that $M_{vac}+\sum_{i=0}^3 M_i=I$. When Eve obtains the outcome corresponding to $M_i$, she resends a standard BB84 state $|\phi_i \rangle$ to Bob. Here we add an additional POVM operator $M_{vac}$, since Eve needs not to resend all the pulses to Bob due to the loss of channel between Alice and Bob. Thus when she obtains the outcome corresponding to this operator, she blocks the pulse and sends a vacuum state to Bob. Obviously, the above attack is the same as the general intercept-and-resend attack. However, in our attack, the dimension of $M_i$ and $M_{vac}$ is three instead of two.

Obviously, Eve can not distinguish the four states $|\Phi_k\rangle$ certainly, since they are linearly dependent states. In the following, we estimate the QBER between Alice and Bob induced by Eve. According to the measurement theory, the conditional probability that Bob obtains $|\phi_j\rangle$ given that Alice sends the state $|\Phi_k\rangle$ is given by:
\begin{equation}
\begin{split}
P(j|k)&=\sum_{i=0}^3 P(B=j|E=i)P(E=i|A=k)\\
&=\sum_{i=0}^3 |\langle \phi_j|\phi_i\rangle|^2Tr(M_i\rho_k),
\end{split}
\end{equation}
where $P(B=j|E=i)$ is the conditional probability that Bob obtains $|\phi_j\rangle$ given that Eve resends the state $|\phi_i\rangle$, $P(E=i|A=k)$ is the conditional probability that Eve obtains outcome corresponding to $M_i$ given that Alice sends $|\Phi_k\rangle$, $\rho_k=|\Phi_k\rangle \langle\Phi_k|$. Note the fact that $|\langle \phi_j|\phi_i\rangle|^2=\delta_{i+2,j}+(\delta_{i+1,j}+\delta_{i+3,j})/2$, where $\delta_{i,j}$ is the kronecker delta function. Therefore, the QBER between Alice and Bob induced by Eve's attack is given by:
\begin{equation}\label{e_Bob}
\begin{split}
e^B&=\frac{\sum_{k=0}^3 \sum_{j=0,j\neq k}^3 P(j|k)}{\sum_{k=0}^3 \sum_{j=0}^3 P(j|k)}\\
&=\frac{\sum_{i=0}^3 Tr(M_i L_i)}{\sum_{i=0}^3 Tr(M_i \rho)}\\
\end{split},
\end{equation}
where
\begin{equation}
\begin{split}
&L_i=\frac{1}{2}\rho_{i+1}+\rho_{i+2}+\frac{1}{2}\rho_{i+3}\\
&\rho=\rho_0+\rho_1+\rho_2+\rho_3.\\
\end{split}
\end{equation}

Generally speaking, the optimal strategy for Eve is to find a set of optimal POVM operators to minimize the QBER, $e_B$. Although Eq.\eqref{e_Bob} is the same as Eq.[3] of Ref.\cite{Fung} formally, the dimension of POVM operators is three in our attack instead of two in Ref.\cite{Fung}. Furthermore, in Ref\cite{Fung}, the authors remark that ``the constraint $\sum_{i=0}^3 M_i\leq I$ is not necessary to minimize $e^B$, since any solution to this unconstrained problem can always be scaled down sufficiently to satisfy the constraint''. This conclusion is also true in our case. However, we remark that although the constraint $\sum_{i=0}^3 M_i\leq I$ will not change the QBER, it will bound the probability that Eve obtains an valid outcome successfully, here the valid outcome means that Eve obtains outcome corresponding to $M_i$ but not $M_{vac}$. The probability that Eve obtains an outcome successfully is also an important parameter to describe Eve's attack, which is given by:
\begin{equation}\label{p_eve}
P_{succ}^E=\frac{1}{4}\sum_{i=0}^3 Tr(M_i\rho),
\end{equation}
where the division by four is that Alice sends one of the four states with equal probability.

Therefore, in order to find the optimal strategy for Eve, she needs to solve the optimization program with two penalty functions, which can be written as:
\begin{equation}
\label{opt_pro}
\begin{matrix}
\text{min} & e^B \\
\text{max} & P_{succ}^E \\
\text{subject to} & M_i\geq 0\\
&M_{vac}\geq 0\\
& M_{vac}+\sum_{i=0}^3 M_i=I\\
\end{matrix}
\end{equation}
where $M_i\geq 0$ and $M_{vac}\geq 0$ mean the matrix $M_i$ and $M_{vac}$ are positive. In fact Eve can not obtain the optimal value of the two penalty functions simultaneous, which will be shown in the following. Thus, in the following, we consider the QBER $e^B$ as the major object and $P_{succ}^E$ as a minor one.

\section{\label{sec:sim}Simulation}
According to the above analysis, the optimal strategy for Eve is given by the solution of Eq.\eqref{opt_pro}. However, it is hard to find the global optimal solution of Eq.\eqref{opt_pro} \cite{Ex_op}. Thus, in this paper, our analysis follows the method of Fung \emph{et al.} \cite{Fung}, who uses this method to analyze the security of QKD system for the phase-remapping attack. In this section, we first introduce the method of Fung and then estimate the QBER between Alice and Bob with numerical simulation. We remark that, although the method considered here is just suboptimal for Eve, it shows clearly that the secrecy of generated key will be compromised by our PFM attack. The numerical simulations show that, under PFM attack, the QBER between Alice and Bob induced by Eve is lesser than 25\% which is the QBER under the general intercept-and-resend attack. Furthermore, when Eve combines our attack with the phase remapping attack, the QBER induced by her attack can be lesser than 11\%, which is the maximal tolerable QBER under the collective attack \cite{Scarani}.

\subsection{The suboptimal strategy for Eve}
According to Ref.\cite{Fung}, the suboptimal strategy can be described as:

\emph{Step one}: Instead of distinguishing the four states sent by Alice, here Eve only distinguishes $\rho_0$ from $\{\rho_1,\rho_2,\rho_3\}$ and $\rho_3$ from $\{\rho_0,\rho_1,\rho_2\}$. It means that, Eve uses three POVM operators $\{M_{vac},M_0,M_3\}$ to measure the states sent by Alice. When she obtains the outcome corresponding to $M_0$ or $M_3$, she resends a standard BB84 state $|\phi_0\rangle$ or $|\phi_3\rangle$ to Bob, otherwise, she resends a vacuum state to Bob. It can also be considered as a special case of the intercept-and-resend attack described in Sec.\ref{sec:attack} that $M_1=M_2=0$.

\emph{ Step two}: Instead of finding the global optimal POVM operators to minimize $e^B$, Eve uses the following POVM operators to measure the states sent by Alice, which are given by:
\begin{equation}\label{povm}
\begin{split}
&M_0=x \rho^{-1/2}|C_0\rangle\langle C_0|\rho^{-1/2}\\
&M_3=x \rho^{-1/2}|C_3\rangle\langle C_3|\rho^{-1/2}\\
\end{split}
\end{equation}
where $|C_0\rangle$ or $|C_3\rangle$ is the eigenvector of matrix $\rho^{-1/2}L_0\rho^{-1/2}$ or $\rho^{-1/2}L_3\rho^{-1/2}$ corresponding to the minimal un-zero eigenvalue $\lambda_0$ or $\lambda_3$, $x$ is the maximal real number that ensure the matrix $M_{vac}=I-M_0-M_3$ is positive. Obviously, $\{M_{vac},M_0,M_3\}$ are positive and sum to the identity, thus they are valid POVM operators.

Here we give two remarks about this suboptimal strategy for Eve.

\emph{Remark one}: Although the global solution of Eq.\eqref{opt_pro} is not obtained in this paper, it does not matter, since our goal is to show the loophole caused by the imperfection of FM, but not to try to give a strict security analysis. Thus if we can find a set of POVM operators that can show this loophole, it is enough. If Eve who has unlimit computation power wants to optimize her strategy, she can solve the optimization program.

\emph{Remark two}: In the suboptimal strategy described above, Bob only obtains bit 0 in one basis and bit 1 in other basis. Glancingly, it will give Alice and Bob a simple method to defeat our attack, since they only need to estimate the count rate of bit 0 and bit 1 for each basis individually. However, we remark that, the method can only defeat our suboptimal strategy, but can not defeat a optimal attack caused by the imperfection of FM. In fact, the suboptimal strategy is just a toy model to show the security loophole induced by the imperfection of FM. This suboptimal strategy is also used by Fung \emph{et al.} in the phase remapping attack to obtain their main conclusion (Fig.(4) of Ref.\cite{Fung}). If a true Eve exists, she can solve the optimal programme Eq.\eqref{opt_pro} and find the optimal strategy for her. Then, the probability that Bob obtains bit 0 and bit 1 is equal in two base.

\subsection{Result}
Substituting the suboptimal POVM operators Eq.\eqref{povm} into Eq.\eqref{e_Bob} and Eq.\eqref{p_eve}, it is easy to estimate the QBER for Bob, $e^B$, and the probability that Eve obtains outcome successfully, $P_{succ}^E$. Since Eve can combines the phase remapping attack \cite{Fung} with our attack, we set the phase difference $\delta \in [0,\pi/2]$ in the following discussion. Here, we remark again that the following results are obtained based on the suboptimal strategy described above. If Eve can find the global solution of Eq.\eqref{opt_pro}, she can improve the following results. Furthermore, note that $\varepsilon=0$ is a singular point in PFM attack, since, in this point, the dimension of Hilbert space spanned by the four states of Eq.\eqref{phi} is reduced to two. It means that Eve can not implement our attack in this point. Thus, this point is excluded in the following simulation.

\begin{figure}
\scalebox{1}{\includegraphics[width=\columnwidth]{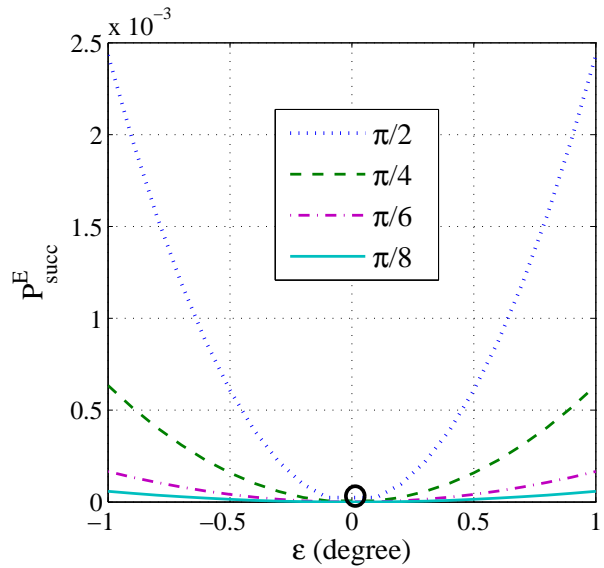}}
\caption{\label{fig:2}(Color online)The probability that Eve obtains outcome successfully changes with $\varepsilon$ for given $\delta$. The fours lines are obtained by maximizing $x$ of Eq.\eqref{povm}. Note that the special point that $\varepsilon=0$ is unconsidered in our simulation. Since this point is a singular point, in which the dimension of Hilbert space spanned by the four states of Eq.\eqref{phi} is reduced to two. Thus our attack is invalid in this point.}
\end{figure}

Fig.\ref{fig:2} shows clearly the probability that Eve obtains outcome successfully, $P_{succ}^E$, changes with $\varepsilon$ for given $\delta$. The larger $\varepsilon$ is, the easier Eve can load her attack. Here, we remark that $P_{succ}^E$ can be explained as the maximal transmittance of channel between Alice and Bob that Eve can load this attack successfully under the suboptimal strategy for given $\varepsilon$ and $\delta$ \cite{Psucc}. For example, when $\varepsilon=1$ and $\delta=\pi/2$, $P_{succ}^E=2.43\times 10^{-3}$. It means that if Eve wants to implement this attack successfully, the transmittance of channel between Alice and Bob should be smaller than $2.43\times 10^{-3}$, which corresponds to a 124km long fiber (the typical loss of fiber is about 0.21 dB/km). Furthermore, $P_{succ}^E$ can also be explained as that, for a given transmittance of channel, Eve can not exploit the imperfection of FM that $\varepsilon$ is smaller than a given value. For example, when $\delta=\pi/2$ and $|\varepsilon|<0.65^\circ$, $P_{succ}^E$ is smaller than $1.029\times 10^{-3}$. Thus for a 142.5km long fiber (the transmittance is about $1.017\times 10^{-3}$), Eve can not exploit the imperfection of FM that $|\varepsilon|<0.65^\circ$.

\begin{figure}
\scalebox{1}{\includegraphics[width=\columnwidth]{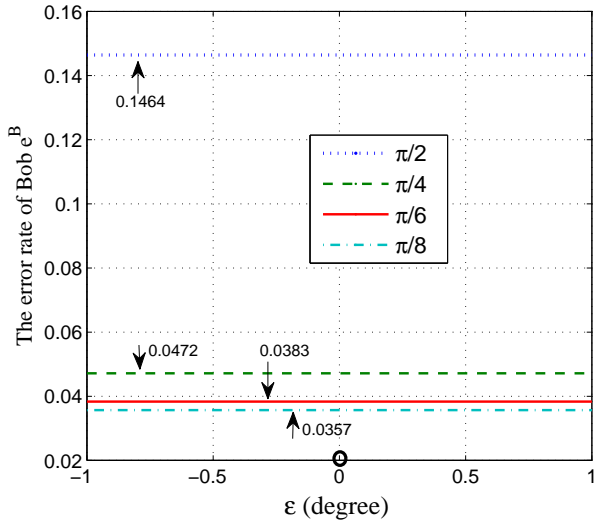}}
\caption{\label{fig:3}(Color online)The error rate of Bob changes with $\varepsilon$ for given $\delta$. Note that, due to the same reason as Fig.\ref{fig:2}, the special point that $\varepsilon=0$ is also unconsidered here. Furthermore, we combine the phase remapping attack with our attack, thus $\delta$ can be smaller than $\pi/2$.}
\end{figure}

The QBER between Alice and Bob induced by Eve's attack is shown in Fig.\ref{fig:3}. It shows that even in the case that $\delta=\pi/2$, the QBER induced by Eve's attack is much lower than 25\% which is QBER induced by the general intercept-and-resend attack. It is also lower than 20\%, which is the maximal tolerable QBER in the BB84 protocol under the two-way post-processing \cite{Chau,Ranade} when Eve does not exploit the imperfection of FM. Furthermore, if Eve combines the phase-remapping attack with our attack, she can reduce the QBER to a very small level. For example, if Eve sets the phase difference $\delta=\pi/8$, the QBER induced by her attack is just 3.57\%, which is lower than 11\% that is the maximal tolerable QBER for the BB84 protocol under the collective attack and one-way post-processing \cite{Scarani}. Then no secret key can be generated when the QBER estimated by Alice and Bob is larger than this value. Therefore, it is necessary for the legitimate parties to consider our attack in a practical \emph{plug-and-play} QKD system.

It is interesting that, when $\delta$ is given, the QBER induced by Eve is almost constant and independent of the degree of the imperfection of FM $\varepsilon$, which is shown in Fig.\ref{fig:3} clearly. The main reason is that $\varepsilon$ is very small. In fact, $e^B$ will changes with $\varepsilon$ slightly, but the difference is so small that it can be ignored. In order to show the conclusion clearly, we consider Eq.\eqref{e_Bob} with $o(\varepsilon^2)$. It is easy to check that, under the suboptimal strategy of Eq.\eqref{povm}, $e^B=(\lambda_0+\lambda_3)/2$, here, $\lambda_0$ and $\lambda_3$ are the minimal un-zero eigenvalue of $\rho^{-1/2}L_0\rho^{-1/2}$ and $\rho^{-1/2}L_3\rho^{-1/2}$. A simple evolution show that, when $\delta=\pi/2$, the three eigenvalues of $\rho^{-1/2}L_0\rho^{-1/2}$ are $1/2$, $(1\pm \sqrt{2}/2)(1-2\varepsilon)/2$. Thus $\lambda_0=(1- \sqrt{2}/2)(1-2\varepsilon)/2=0.1464-0.2929\varepsilon$. The same result can be obtained for $\lambda_3$. Therefore, the error rate of Bob induced by Eve can be written as $e^B=0.1464+o(\varepsilon)$, which shows clearly that $e^B$ is constant in order of $o(\varepsilon)$. In fact, the strict numerical simulation shows that, for given $\delta$, $e^B$ is almost constant in order of $10^{-5}$ for each $\varepsilon$. Note that, although $e^B$ is almost independent of $\varepsilon$, $\varepsilon$ will affect $P_{succ}^E$ obviously (see Fig.\ref{fig:2}).

Fig.\ref{fig:3} shows that when Eve combines the phase remapping attack with the imperfection of FM, the QBER induced by her attack can be reduced to a very small level. For example, when she sets $\delta=\pi/4$, the QEER is just 4.72\%. However, when the QBER is reduced, the probability that Eve implements her attack successfully will also be reduced, see Fig.\ref{fig:2}. Therefore, when Eve loads her attack, she should make a trade-off between the QBER and the efficiency to maximize her information.

\begin{figure}
\scalebox{1}{\includegraphics[width=\columnwidth]{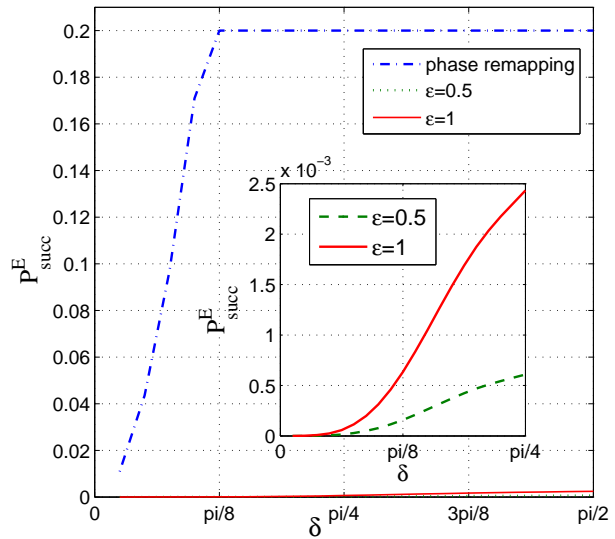}}
\caption{\label{fig:4}(Color online) The probability that Eve obtains outcome successfully changes with $\delta$ for the phase remapping attack and our attack. The red line is obtained for the phase remapping attack according to the method described in Ref.\cite{Fung}. The blue and green lines are obtained for $\varepsilon=1^\circ$ and $\varepsilon=0.5^\circ$, respectively.}
\end{figure}

\begin{figure}
\scalebox{1}{\includegraphics[width=\columnwidth]{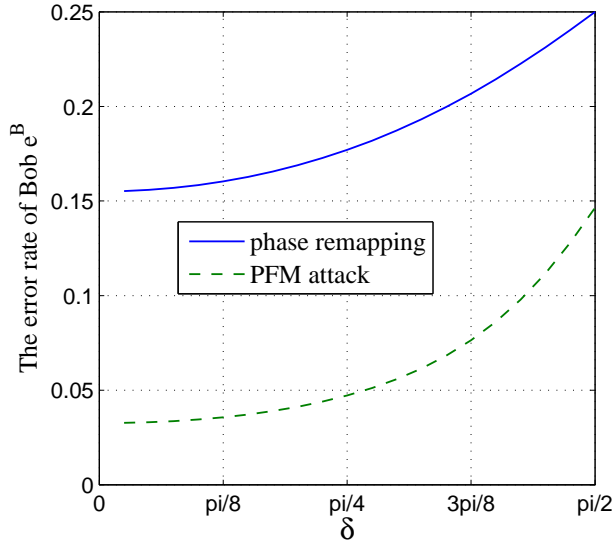}}
\caption{\label{fig:5}(Color online)The error rate of Bob changes with $\delta$ for the phase remapping attack and PFM attack. In the simulation, we set $\varepsilon=1^\circ$.}
\end{figure}

In the following, we compare our attack with the phase remapping attack \cite{Fung}. The probability that Eve obtains outcome successfully is shown in Fig.\ref{fig:4}. It shows that the probability that Eve implements the phase remapping attack successfully is much larger than that of our attack. Although Eve can increase the probability that she implements her attack successfully by increasing the phase difference $\delta$, it will induce more QBER which is shown in Fig.\ref{fig:5}.

Fig.\ref{fig:5} shows that, when the phase difference $\delta$ is increased, the QBER induced by Eve's attack will increase quickly. This conclusion holds for both the phase remapping attack and our attack. However, the QBER under our attack is much lower than that of the phase remapping attack. For example, when $\delta=\pi/4$, $e^B=17.7\%$ for the phase remapping attack, but in our attack, $e^B=4.71\%$.

\section{\label{sec:con}CONCLUSION}
In the two-way \emph{plug-and-play} QKD system, a perfect $90^\circ$ FM can be used to compensate the birefringence of fiber automatically and perfectly. However, the practical FM is imperfect. Although the deviation from the ideal case is small and the QBER induced by this imperfection is slight, it will leave a loophole for Eve to spy the secret key between Alice and Bob. In fact, when the practical FM deviates from the ideal case, the dimension of Hilbert space spanned by the states sent by Alice is three instead of two. Thus the standard security analysis is invalid here, some careful strategy should be adopted by the legitimate parties to monitor this imperfection.

In this paper, we propose a PFM attack in two way \emph{plug-and-play} QKD system based on the imperfection of FM. The results show that, under this attack, the QBER between Alice and Bob induced by Eve is much lower than 25\%, which is the QBER for the general intercept-and-resend attack when FM is perfect. Furthermore, when Eve combines the imperfection of FM with phase remapping attack, the QBER induced by her attack can be lower than 11\%, which is the maximal tolerable QBER for the BB84 protocol under the collective attack and one-way post-processing. Therefore, in the practical case, the legitimate parties should pay more attention to the imperfection of FM, otherwise, the secrecy of generated key will be compromised. However, we remark that, although Eve can combine PFM attack with phase-remapping attack to reduce QBER between Alice and Bob induced by her attack, the probability that she can load this attack successfully is dependent on the loss of channel obviously. In other words, Eve can only implement PFM attack in long distance QKD system, which can be estimated in Fig.\ref{fig:2} and Fig.\ref{fig:4}.

\section{ACKNOWLEDGEMENT}
This work is supported by National Natural Science Foundation of China Grants No.61072071. Shi-Hai Sun is supported by Hunan Provincial Innovation Foundation For Postgraduate No.CX2010B007 and Fund of Innovation, Graduate School of NUDT No.B100203.



\begin{thebibliography}{100}
\bibitem{Ex_op} There are three problems to find the global optimal solution of Eq.(14): (1) There are two penalty functions, it is needed to make a trade-off between them, thus it is a problem to deal with it in the numerical simulation. (2) For the five POVM operators belonging to 3-D Hilbert space, the number of variable is 39, it is too much for the simulation. (3)In our problem, the geometrical format of the restrictions is irregular, thus it is hard to find a suitable algorithm to solve this problem. However, it does not matter, since we just want to show the loophole caused by the imperfection of FM, if we can find a set of POVM operators to show this loophole, it is enough.
\bibitem{Psucc} Here we associate $P_{succ}^E$ with the channel transmittance, but do not consider the transmittance of Bob's setups which includes both the transmittance of Bob's optical setups and the efficiency of Bob's single photon detector (SPD), there are two reasons for this consideration: (1) The channel can be controlled by Eve, but the Bob's setups are placed in Bob's secure zone. Thus Eve can replace the practical channel with a lossless channel, but she cannot change the efficiency of Bob's detectors. (2) In order not to be discovered, Eve should keep Bob's count rate unchanged. When Eve is absent, Bob's count rate is $C \propto \eta_c \eta_b$, where $\eta_c$ is the channel transmittance between Alice and Bob, $\eta_b$ is the transmittance of Bob's setups. In this paper, we assume Eve resends a single photon pulse to Bob according to her detection result. Thus $C\propto P_{succ}^E \eta_b$, which means $P_{succ}^E=\eta_c$. In fact, we remark that, although Eve cannot change the transmittance of Bob's setups, she can compensate it with a birght light whose intensity is $\mu$. Then $C\propto \mu P_{succ}^E \eta_b$. Thus, when $\mu=1/\eta_b$, $P_{succ}^E=\eta_c\eta_b$, which means Eve also can associate $P_{succ}^E$ with the overall efficiency of QKD system by compensating the transmittance of Bob's setups. However, when a birght light is used by Eve, it will introduce other problems, for example, the double click. Therefore, in this paper, we only associate $P_{succ}^E$ with the channel transmittance.
\end{thebibliography}
\end{document}